\documentclass[pre,showkeys]{revtex4}

\usepackage{amsmath}
\usepackage{amsfonts}
\usepackage{bm}
\usepackage{epsfig}

\begin{document}

\title{Regular binary thermal lattice-gases}

\author{Ronald Blaak}
\thanks{Current Address: Institut f\"ur Theoretische Physik II,
Heinrich-Heine-Universit\"at, D-40225 D\"usseldorf, Germany;
e-mail:blaak@thphy.uni-duesseldorf.de} 
\affiliation{Grupo Interdisciplinar de Sistemas Complicados (GISC),
Departamento de Matem\'aticas, Universidad Carlos III de Madrid,
Avda. de la Universidad, 30, 28911, Legan\'es, Madrid, Spain}

\author{David Dubbeldam}
\email{dubbelda@science.uva.nl}
\affiliation{Department of Chemical Engineering, University of
Amsterdam, Nieuwe Achtergracht 166, 1018 WV Amsterdam, The
Netherlands}

\begin{abstract}
We analyze the power spectrum of a regular binary thermal lattice
gas in two dimensions and derive a Landau-Placzek formula, describing
the power spectrum in the low-wavelength, low frequency domain, for
both the full mixture and a single component in the binary
mixture. The theoretical results are compared with simulations
performed on this model and show a perfect agreement. The power 
spectrums are found to be similar in structure as the ones obtained
for the continuous theory, in which the central peak is a complicated
superposition of entropy and concentration contributions, due to the
coupling of the fluctuations in these quantities. Spectra based on the
relative difference between both components have in general additional
Brillouin peaks as a consequence of the  equipartition failure.
\end{abstract}


\maketitle

\section{Introduction}
Lattice-gas automata (LGA), as introduced by  Frisch, Hasslacher, and
Pomeau \cite{Fri86}, are ideal testing systems for kinetic theory.
Although they have a simple structure, which makes them extremely
efficient simulation tools, they still address the full many-body
particle problem.
Much of the efficiency originates from the discretization of the
positions and velocities of the point-like particles onto a lattice. 
The simplified
dynamics is a cyclic process: a streaming step, where all particles propagate 
to neighboring lattice sites, followed by a local collision step.
The collisions typically conserve mass and momentum, and in addition
energy if the model under consideration is thermal.

Lattice-gases are capable of simulating macroscopic fluid
flow~\cite{Fri86,Boon2001}, and can be used
for studying flow through porous media~\cite{Chen}, immiscible 
multicomponent fluids~\cite{Gunstensen},
reaction-diffusion~\cite{ReactionDiffusion}, von Karman
streets~\cite{vonKarmanstreet}, Rayleigh-B\'enard
convection~\cite{RayBer}, interfaces and phase
transitions~\cite{Rothman:1994RMP}.
However, there are practical problems when using LGA's: 
the models are not Galilean-invariant, temperature is not well-defined,
the transport coefficient have unexpected behavior as a function of
temperature and density, and many models contain spurious
invariants~\cite{Brito:1991JSP,Das:1992PA,Ernst:1992JSP}. It is 
therefore difficult to make connections with realistic systems.
Although in the early years the LGA's were thought to seriously compete
with other fluid flow solvers, the main line of research has shifted to
a testing ground for concepts in kinetic theory. The main motivation for
this paper lies in the exploration of the limits of thermal
lattice-gas capabilities concerning diffusion-like phenomena.

Diffusion can be incorporated in LGA's in several ways. The
computational most efficient method is the \emph{color
mixture}~\cite{Boon2001,Ernst:1990JSP,Hanon}, which we have analyzed
in detail for a thermal model~\cite{Blaak:2001PRE}. The otherwise
identical particles are in such a mixture painted with a probability 
corresponding to the concentration and a color-blind observer would
not notice any difference, i.e. all transport properties are the same as
for the single-component fluid. But in addition there is an extra
hydrodynamical and color-dependent diffusion mode that does not
interfere with the other modes.  

In this paper we consider a different way to include diffusion,
namely the regular binary mixture~\cite{Ernst:1990JSP}. In the regular
binary mixture one has two distinct species of particles,
i.e. different mass. For particles of the same species the exclusion
principle holds and hence there can be at most one particle of a given
species in any velocity channel. There is, however, no mutual
exclusion for particles off different species. Consequently, a given
velocity channel can be occupied by two particles, provided they both
belong to a different species.

One can interpret this as each species living on its own but identical
lattice with exclusion. Since each specie is restricted to its own
lattice there can be no mass exchange between the two lattices and
there is local mass conservation for both species. Upon interaction,
however, particles of both lattices corresponding to the same node are
able to exchange momentum and energy, provided this does no violate
local mass conservation and exclusion for each species. Consequently,
the dynamics of the regular binary mixture is much more involved than
that of the color mixture, but it is closer to the dynamics of real
fluids. For simplicity we here will assume the special case of the two
species having the same physical properties, i.e. the same mass, and
distinguish both species by a different color label. 

We analyze the model at the microscopic level and focus mainly on the
behavior of the power spectrum. The coupling between energy transport and 
diffusion that arises in these systems, results in a more complicated
structure of the power spectrum
\cite{Mountain,BoonYip,Blaak:2001PREb}, in which the  
central peak now contains combined effects of both entropy
fluctuations and concentration fluctuations. Macroscopically this
coupling manifests itself as the Dufour effect (a concentration
gradient induces a heat-flow) and the Soret effect (a temperature
gradient induces a diffusion flux).  

The remainder of this paper is organized as follows. We start by
introducing the regular binary thermal lattice gas model in section
\ref{Sec:Model} and use the molecular chaos assumption to obtain the
linearized collision operator. Section \ref{Sec:Mode Analysis} is
concerned with a mode analysis of the linearized system,
revealing the appearance of an extra diffusion mode. The modes related to
thermal diffusivity and mass-diffusion are coupled and form two
non-propagating, diffusive modes. Furthermore, in section
\ref{Sec:Landau-Placzek} we derive a Landau-Placzek analogue, a
formula which describes the power spectrum in the hydrodynamic limit
(long wavelength, small frequency domain), which we check with
simulation results performed on this system in section
\ref{Sec:simulation}. In the final section we give a brief overview of
the main results and some concluding remarks.

\section{The regular binary GBL-model}
\label{Sec:Model}

\begin{center}
\begin{figure}
\epsfig{figure=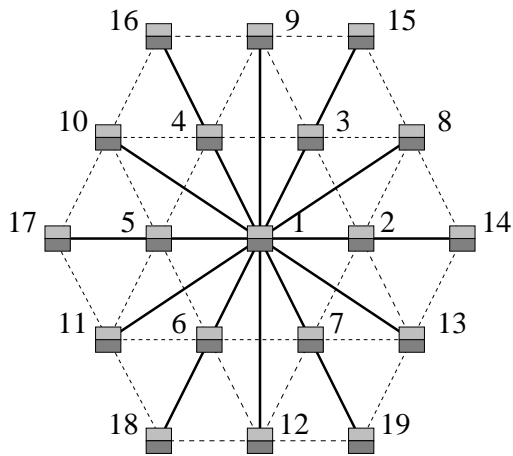,width=16pc,angle=0}
\caption{\label{Fig:spatial layout} 
The spatial layout of the velocity set of the regular binary
mixture. It is identical to the one for the GBL model, but now a
channel can be occupied by a red, a blue or both at the same time.}
\end{figure}
\end{center}

The LGA model we like to consider here is a thermal model consisting
of two different interacting species of particles with identical
mass. They can be thought 
of to live on separate two-dimensional hexagonal lattices on which they
propagate independently, but red and blue particles corresponding to
the same node 
interact during the collision. The velocities are discretized and have 
a spatial layout as  shown in figure \ref{Fig:spatial layout}. They are
distributed over four rings: one ring contains a single null-velocity,
and three rings each contain six velocities with magnitude $1$,
$\sqrt{3}$, and $2$.
The velocity set is of size $19$, and equivalent to the GBL-model, proposed
by Grosfils, Boon, and Lallemand~\cite{GBL92}. 
Since the model under consideration is basically a combination of two
GBL models, particles of different type (red and blue)
can have the same velocity. However, there can not be more than a
single particle of each type in a given velocity state. The multiple
rings correspond each to a different energy level and are a necessary
requirement in order to introduce thermal properties. This particular
velocity set guarantees the absence of spurious invariants and results in
macroscopically isotropic behavior \cite{GBL93}. 

The state of a node can be specified by a set of boolean occupation
numbers $n_{i\mu}$, denoting the  presence or absence of a particle of
type $\mu=\{r,b\}$ in velocity channel $\bm{c}_i$, where $i$ is a label
running over all 19 velocities. Due to the boolean nature of the LGA,
the ensemble average of the occupation numbers $f_{i\mu}$ in
equilibrium, is described by a Fermi-Dirac distribution \cite{Boon2001}
\begin{equation}
\label{Eq:FermiDirac}
f_{i\mu} \equiv \langle n_{i\mu}\rangle = \frac{1}{1 + e^{- \alpha_\mu
+ \frac{1}{2} \beta \bm{c}_i^2 -\bm{\gamma} \cdot \bm{c}_i}},  
\end{equation}
where $\alpha_r$, $\alpha_b$, $\beta$, and $\bm{\gamma}$ are
Lagrange multipliers and fixed by setting the value of the average
red density $\rho_r=\sum_{ir} f_{ir}$, the average blue density
$\rho_b=\sum_{ib} f_{ib}$ (or alternatively the total density
$\rho=\rho_r + \rho_b$ and the fraction $P_r$ of red particles), the
average velocity $\rho\bm{ u}=\sum_{i\mu} f_{i\mu}\bm{ c}_i$,
 and the energy density $\rho e=\frac{1}{2}\sum_{i\mu} f_{i\mu}\bm{
c}_i^2$. Here, $\beta$ is the inverse temperature, $\alpha_r$ and
$\alpha_b$ fulfill a chemical potential role, and
$\bm{ \gamma}$ is a parameter conjugate to the flow velocity. In the
remainder of this paper we will restrict ourselves to the
global zero-momentum case by putting $\bm{ \gamma}=0$.

The lattice-gas Boltzmann equation, describing the time-evolution of the
average occupation numbers, is given by~\cite{Boon2001}
\begin{equation}
 f_{i\mu}(\bm{ r}+\bm{ c}_i,t+1)=f_{i\mu}(\bm{ r},t)+\Delta_{i\mu}(f),
 \label{Eq:NonlinearBoltzmannequation}
\end{equation}
where the nonlinear collision term $\Delta_{i\mu}(f)$ is a summation over all
pre-and post-collision states $s$ and $s'$
\begin{equation}
 \Delta_{i\mu}(f)=\sum_{s,s'}P(s) A(s\rightarrow s') (s_{i\mu}'-s_{i\mu}).
\end{equation}
The collision rules that are used are taken in to account by the transition
matrix $A$, which contains the probability that on collision a state
is transformed into an ``equivalent''   
state. We denote the collection of states that can be transformed into
each other by an equivalence class $\Gamma=(M_r,M_b,\bm{ P},E)$,
i.e. a class having the same red mass
$M_r=\sum_i s_{ir}$, the same blue mass $M_b=\sum_i s_{ib}$, the same
momentum $\bm{P}=\sum_{i\alpha} s_{i\alpha}\bm{ c}_i$, and the same
energy $E=\sum_{i\alpha}\frac{1}{2}s_{i\alpha}\bm{ c}_i^2$.
We adopt maximal collision rules, i.e. under collision a state can
transform in any other state within the same equivalence class,
including itself, thus fulfilling all conservation laws. Hence
$A(\Gamma)=\frac{1}{|\Gamma|}$, where we 
use $|\Gamma|$ to denote the number of elements in the class $\Gamma$.
The density and energy density enter the collision operator through $P(s)$,
the probability of occurrence of state $s$ in equilibrium. In the Boltzmann
approximation, the velocity channels are assumed to be independent,
hence $P(s)$ can be written as 
\begin{equation}
  P(s)=\prod_{i\mu} f_{i\mu}^{s_{i\mu}} (1-f_{i\mu})^{1-s_{i\mu}}.
\end{equation}

If the velocity fluctuations are sufficiently small a Taylor expansion of
the collision term in the neighborhood of the equilibrium distribution is
justified~\cite{Boon2001}, yielding the linearized collision operator $\Omega$
\begin{equation}
\label{Eq:OmegaKappa}
 (\Omega\kappa)_{i\mu,j\nu}=\sum_{s,s'} P(s)A(s\rightarrow s')
    (s_{i\mu}'-s_{i\mu})s_{j\nu},
\end{equation}
where the diagonal matrix $\kappa_{i\mu,j\nu}$ is determined by
$\kappa_{i\mu,i\mu}=f_{i\mu}(1-f_{i\mu})$, which is the variance
in the occupation number of a channel with the corresponding
labels. It can easily be checked from Eq.~(\ref{Eq:OmegaKappa}) that 
$\Omega\kappa$ is a symmetric $38\times38$ matrix. In contrast with
the colored LGA \cite{Blaak:2001PRE}, $\kappa$ is a 38-dimensional
diagonal matrix due to the fact that there is no mutual exclusion of a
red and blue particle with the same velocity. It also enables us to 
introduce the colored thermal scalar product
\cite{Hanon,Blaak:2001PRE,Ernst:1992JSP}, 
\begin{align}
 \label{Eq:inproduct}
 \langle A|B\rangle&=\sum_{i\mu}A(\bm{ c}_{i\mu})
 B({\mathbf c}_{i\mu}) \kappa_{i\mu},\\
 \langle A|\Omega|B\rangle&=\sum_{i\mu j\nu}A(\bm{ c}_{i\mu})
    (\Omega\kappa)_{i\mu,j\nu} B(\bm{ c}_{j\nu}),
 \label{Eq:thermal inner matrix}
\end{align}
where we adopt the convention that the matrix $\kappa$ is attached to
the right vector.

\section{Perturbation theory}
\label{Sec:Mode Analysis}

In first approximation the behavior of the system can be obtained by 
considering the fluctuations $\delta f_{i\nu}(\bm{ r},t) = f_{i\nu}
(\bm{ r},t)-f_{i\nu}$ and analyze these deviations from the uniform
equilibrium state in terms of eigenmodes. The solutions of the
Boltzmann equation (\ref{Eq:NonlinearBoltzmannequation}) are then of
the form 
\begin{equation}
\delta f_{i\nu} = \psi_\mu(\bm{ k,c_i}) e^{-\imath \bm{ k} \cdot
\bm{ r}+ z_\mu(\bm{ k})t}, 
\end{equation}
where $z_\mu(\bm{ k})$ represents the eigenvalue of the mode
$\psi_\mu$ at wavevector $\bm{ k}$. The hydrodynamic modes are
related to the collisional invariants and satisfy $z_\mu\rightarrow 0$
for $\bm{ k}\rightarrow 0$. There are five independent collision
invariants $a_n$  corresponding to the conservation of the red and 
blue mass, the conservation of momentum, and the conservation of
energy
\begin{equation}
\label{Eq:Invariantequation}
\langle a_n| \Omega = 0 \qquad \Omega | a_n \rangle = 0,
\end{equation}
\begin{equation}
\label{Eq:Invariants}
|a_n \rangle = \{ |R\rangle, |B\rangle, |c_x\rangle, |c_y\rangle,
 |\frac{1}{2} c^2\rangle\},
\end{equation}
where $|R\rangle_{i\alpha} = \delta_{\alpha r}$, $|B\rangle_{i\alpha}
= \delta_{\alpha b}$, and can be combined to form $|\rho\rangle =
|R\rangle + |B\rangle$.  
It turns out that an equivalent and more convenient set of invariants
is given by the following five combinations which are mutual orthogonal
with respect to the colored thermal scalar product (\ref{Eq:inproduct})
\begin{equation}
\label{Eq:Invariantbasis}
 |a_n \rangle =\{ |c_x\rangle, |c_y\rangle, 
 |p\rangle = |\frac{1}{2} c^2\rangle, |s\rangle = |p\rangle - c_s^2
 |\rho\rangle, 
 |d\rangle\},
\end{equation}
where $|s\rangle$ is the analogue to the microscopic entropy for the
GBL model~\cite{GBL93} and $c_s^2=\frac{\langle p|p\rangle}{\langle
p|\rho\rangle}$ the speed of sound following from the orthogonality
requirement $\langle p|s\rangle = 0$. The remaining invariant
$|d\rangle$ is, contrary to the one in the colored GBL model
\cite{Blaak:2001PRE}, not simply the weighted difference between the
red and blue densities $\frac{1}{\langle R|R\rangle}|R\rangle-
\frac{1}{\langle B|B\rangle}|B\rangle$, but is given by
\begin{equation}
\label{Eq:invariant-d}
|d\rangle = \left( \frac{|R\rangle}{\langle s|R\rangle} -
\frac{|B\rangle}{\langle s|B\rangle}\right) - 
\left( \frac{\langle p|R\rangle}{\langle s|R\rangle} -
\frac{\langle p|B\rangle}{\langle s|B\rangle}\right) 
\frac{|p\rangle}{\langle p|p\rangle}.
\end{equation}
This form is determined by the requirement that it is orthogonal to
the four other invariants. In the case of equal red and blue density
the last term on the righthand side vanishes, but in general this is
not the case. The origin of this term is in fact the absence of
equipartition in this model due to the exclusion principle, which
causes the ratio $f_{ir}/f_{ib}$ to be dependent on the velocity of
the particles as can be seen from Eq.~(\ref{Eq:FermiDirac}). 
Note that this last invariant is also perpendicular to the density,
i.e. $\langle \rho|d\rangle = 0$. Other useful relations between
these invariants are $\langle \rho|p\rangle = \langle \rho|c_x^2
\rangle = \langle \rho|c_y^2\rangle$  and $\langle p|p \rangle =
\langle p|c_x^2 \rangle = \langle p|c_y^2\rangle$.

The choice of $|s\rangle$ and $|d\rangle$ made here is based on the
correspondence of the definition of $|s\rangle$ with the one made for
the GBL model and its colored counterpart
\cite{Blaak:2001PRE,GBL93}. Since $|s\rangle$ nor $|d\rangle$, with
the exception of some special limits, is going to be the zeroth
order of an eigenmode the choice is somewhat arbitrary and any two
linear combinations that are mutual orthogonal could be used as well.
The current choice, however, will facilitate us to make a connection
with continuous theory and the proper transport coefficients.

Note that this is similar to what is done in the case of continuous
theory and generalized versions of hydrodynamics (See
Ref. \cite{BoonYip} and references therein), where one also 
needs to obtain a set of variables that are statistically
independent based on thermodynamic fluctuation theory. This does not
uniquely fix these variables, and the freedom that remains can be used
in order to select an appropriate, orthogonal set for the specific
problem.  

Following the method of R\'esibois and Leener \cite{Resibois} we need
to find the $\bm{ k}$-dependent eigenfunctions and eigenvalues of
the single-time step Boltzmann propagator
\begin{equation}
\label{Eq:eigenproblem-right}
e^{-\imath \bm{k}\cdot \bm{c}}(\bm{1}+\Omega)
|\psi(\bm{k})\rangle = e^{z(\bm{k})}| \psi(\bm{k})\rangle,
\end{equation}
\begin{equation}
\label{Eq:eigenproblem-left}
\langle \phi(\bm{k})| e^{-\imath \bm{k}\cdot
\bm{c}}(\bm{1}+\Omega) = e^{z(\bm{k})} \langle \phi(\bm{k})|,
\end{equation}
where $e^{-\imath \bm{k} \cdot \bm{c}}$ has to be interpreted as a
diagonal matrix, and $\bm{1}$ is the identity matrix. The symmetries
of the matrices cause the left and right eigenvectors to be related by
$\phi_\mu({\bf k})=e^{\imath \bm{k}\cdot \bm{c}} \psi_\mu({\bf
k})/{\cal M}_\mu$, and form a complete biorthonormal set
\begin{equation}
\label{Eq:biortho}
\sum_\mu |\psi_\mu\rangle\langle\phi_\mu| = \bm{1} \qquad \langle\phi_\mu |
\psi_\nu\rangle = \delta_{\mu\nu},
\end{equation}
where we used $\mu$ and $\nu$ to label the different eigenfunctions and
introduced the normalization constants ${\cal M}_\mu$.

To obtain the hydrodynamic modes characterized by $z(\bm{k})
\rightarrow 0$ in the limit $\bm{k} \rightarrow 0$, we make a Taylor 
expansion of the eigenfunctions and eigenvalues
\begin{equation}
\psi_\mu(\bm{k}) = \psi_\mu^{(0)} + (\imath k) \psi_\mu^{(1)} +
(\imath k)^2 \psi_\mu^{(2)} + \cdots,
\end{equation}
\begin{equation}
z_\mu(\bm{k}) = (\imath k) z_\mu^{(1)} + (\imath k)^2 z_\mu^{(2)} + \cdots,
\end{equation}
where we already used that $z_\mu^{(0)}=0$. The functions
$\psi_\mu(\bm{k})$ will be determined up to a normalization
factor. As this normalization is not allowed to be observable, and in
fact will not be observable, we can choose it in a convenient way by
\begin{equation}
\langle \psi_\mu^{(0)}|\psi_\mu(\bm{k})\rangle =
\langle \psi_\mu^{(0)}|\psi_\mu^{(0)}\rangle,
\end{equation}
which leads to
\begin{equation}
\label{Eq:normalisation}
\langle \psi_\mu^{(0)}|\psi_\mu^{(n)}\rangle =
\delta_{ n 0} \langle \psi_\mu^{(0)}|\psi_\mu^{(0)}\rangle.
\end{equation}
Substitution of the expansions in Eq.~(\ref{Eq:eigenproblem-right}) and
grouping according to the same order in $k$ gives
\begin{equation}
\Omega |\psi_\mu^{(0)}\rangle = 0,
\end{equation}
\begin{equation}
\label{Eq:order-one}
\Omega |\psi_\mu^{(1)}\rangle = (c_\ell + z_\mu^{(1)})
|\psi_\mu^{(0)}\rangle,
\end{equation}
\begin{equation}
\label{Eq:order-two}
\Omega |\psi_\mu^{(2)}\rangle = (c_\ell + z_\mu^{(1)})
|\psi_\mu^{(1)}\rangle + \left[ z_\mu^{(2)} + \frac{1}{2} (c_\ell +
z_\mu^{(1)} )^2 \right] |\psi_\mu^{(0)}\rangle,
\end{equation}
with $c_\ell = \hat{\bm{k}} \cdot \bm{c}$ and its
orthogonal counterpart
$c_\perp = \hat{\bm{k}}_\perp \cdot \bm{c}$. 

The solution of the zeroth order equation is straightforward and 
gives
\begin{equation}
|\psi_\mu^{(0)}\rangle = \sum_n A_{\mu n} |a_n\rangle,
\end{equation}
with some unknown coefficients $A_{\mu n}$. Substitution in
the first order equation and multiplying with $\langle a_m|$ at the
left gives for each $m$ a linear equation in the $A_{\mu n}$
\begin{equation}
\sum_n A_{\mu n} \langle a_m|c_\ell + z_\mu^{(1)} |a_n\rangle = 0,
\end{equation}
where we used that $\langle a_m|\Omega = 0$. This set of equations has
a solution provided the determinant is zero, leading to the following
eigenvalue equation for $z_\mu^{(1)}$
\begin{equation}
\left(z_\mu^{(1)}\right)^3 \left( \langle p|p\rangle \langle
c_\ell|c_\ell\rangle \left(z_\mu^{(1)}\right)^2 -  \langle
p|c_\ell\rangle^2 \right) = 0,
\end{equation}
where it is used that combinations containing an odd power of $c_\ell$
or $c_\perp$ are necessarily zero. This determines two of the five
zeroth order eigenfunctions
\begin{equation}
\label{Eq:soundmodes}
|\psi_\sigma^{(0)}\rangle = |p\rangle +\sigma c_s |c_\ell\rangle \qquad
 z_\pm^{(1)} = - \sigma c_s,
\end{equation}
where $\sigma=\pm$ denotes opposite directions parallel to $\bm{ k}$ and
\begin{equation}
 \left(z_\sigma^{(1)}\right)^2=\frac{\langle p|c_\ell^2\rangle^2}
  {\langle p|p\rangle \langle p|\rho\rangle}=c_s^2.
\end{equation}
For the remaining three eigenfunctions we can only conclude at this
level that they satisfy $z_\mu^{(1)} = 0$ and are formed by
combinations of $|d\rangle$, $|s\rangle$, and $|c_\perp\rangle$
only. On physical grounds it can be argued that $|c_\perp\rangle$
will be a mode on itself, however, it also will follow in a natural
way later on.

The combination $ |j_\mu\rangle = c_\ell + z_\mu^{(1)}
|\psi_\mu^{(0)}\rangle$ is called the current. We will extend the
definitions of the current $j_\mu$ and $\psi_\mu^{(0)}$ to include $s$
and $d$ as a possible value for $\mu$. This is for convenience, since
strictly speaking $|s\rangle$ and $|d\rangle$ will in general not be
the source of an eigenfunction, but both will be a linear combination
of two ``true'' modes of the system.

With the introduction of these currents it follows immediately from
Eqns.~(\ref{Eq:Invariantequation}) and (\ref{Eq:order-one}) that 
invariants and currents are orthogonal
\begin{equation}
\label{Eq:psi0-j}
\langle\psi_\nu^{(0)}|j_\mu\rangle = 0,
\end{equation}
hence the currents lie in the complement of the null-space. Since this
is not affected by applying $\Omega$ we obtain
\begin{equation}
\label{Eq:psi0-j-m}
\langle\psi_\nu^{(0)}|\Omega^m|j_\mu\rangle = 0
\end{equation}
for any integer $m$, including negative values. Note that for negative
values of $m$ the expression $\Omega^m|j_\mu\rangle$ is uniquely
determined by the requirement that the result has no component in the
null-space. 
Therefore the solution of the first order equation
(\ref{Eq:order-one}) can formally be written as
\begin{equation}
\label{Eq:solution-one}
|\psi_\mu^{(1)}\rangle = \frac{1}{\Omega}|j_\mu\rangle + \sum_\nu
 B_{\mu \nu} |\psi_\nu^{(0)}\rangle,
\end{equation}
with yet to be determined coefficients $B_{\mu \nu}$ and also three
still unknown currents. The coefficients $B_{\mu\mu}=0$ as
follows from the chosen normalization (\ref{Eq:normalisation}) of the
eigenfunctions. 

Multiplying the second order equation (\ref{Eq:order-two}) with
$\langle\psi_\lambda^{(0)}|$ at the left side we find
\begin{eqnarray}
0 = & \langle\psi_\lambda^{(0)}| (c_\ell + z_\lambda^{(1)})
|\psi_\mu^{(1)}\rangle + (z_\mu^{(1)} - z_\lambda^{(1)})
\langle\psi_\lambda^{(0)}|\psi_\mu^{(1)}\rangle + \nonumber\\ & 
z_\mu^{(2)} \langle\psi_\lambda^{(0)}|\psi_\mu^{(0)}\rangle + \frac{1}{2}
\langle\psi_\lambda^{(0)}|(c_\ell + z_\lambda^{(1)}) (c_\ell +
z_\mu^{(1)}) |\psi_\mu^{(0)}\rangle + \nonumber\\ & \frac{1}{2} (z_\mu^{(1)} -
z_\lambda^{(1)}) \langle\psi_\lambda^{(0)}|(c_\ell + z_\mu^{(1)})
|\psi_\mu^{(0)}\rangle.
\end{eqnarray}
Using the definition of the currents, substitution of the formal
solution of the first order equation (\ref{Eq:solution-one}),
and using the orthogonality relations (\ref{Eq:psi0-j}) this leads to
\begin{equation}
\label{Eq:second-order}
z_\mu^{(2)} \langle\psi_\lambda^{(0)}|\psi_\mu^{(0)}\rangle +
(z_\mu^{(1)} - z_\lambda^{(1)}) B_{\mu \lambda} \langle
\psi_\lambda^{(0)} |\psi_\lambda^{(0)}\rangle = - \langle j_\lambda |
\frac{1}{\Omega} + \frac{1}{2}| j_\mu\rangle.
\end{equation}
For $\lambda=\mu$ this gives the five transport coefficients
\begin{equation}
\label{Eq:transport}
z_\mu^{(2)} = - \frac{\langle j_\mu | \frac{1}{\Omega} +
\frac{1}{2}| j_\mu\rangle}{\langle \psi_\mu^{(0)} | \psi_\mu^{(0)}
\rangle},
\end{equation}
and for $\lambda \neq \mu$
\begin{equation}
\label{Eq:B-mu-nu}
(z_\mu^{(1)} - z_\lambda^{(1)}) B_{\mu \lambda} \langle
\psi_\lambda^{(0)} |\psi_\lambda^{(0)}\rangle = - \langle j_\lambda |
\frac{1}{\Omega} + \frac{1}{2}| j_\mu\rangle,
\end{equation}
from which some of the values of $ B_{\mu \lambda}$ can be obtained,
provided that $(z_\mu^{(1)} - z_\lambda^{(1)})$ is nonzero, and of
which the resulting expressions can be found in
Appendix~\ref{Sec:appendix}. Note that at this stage three of the
zeroth order eigenfunctions and their corresponding currents are still
undetermined. 

In the space of $|d\rangle$, $|s\rangle$, and $|c_\perp\rangle$ one
can rewrite Eq.~(\ref{Eq:second-order}) in the form of a new
eigenvalue problem in the still undetermined coefficients $A_{\mu n}$
\begin{equation}
\sum_n \left( \langle j_m | \frac{1}{\Omega} + \frac{1}{2}| j_n\rangle
 + z_\mu^{(2)} \langle a_m | a_n \rangle \right) A_{\mu n} = 0.
\end{equation}
Equating the determinant to zero we find that there are three different
eigenvalues and we can obtain the form of the remaining three zeroth
order eigenfunctions. Using the fact that there is no degeneracy we know
that we can diagonalize the matrix in terms of the proper
functions. Hence the off-diagonal matrix elements $\langle j_\mu |
\frac{1}{\Omega} + \frac{1}{2}| j_\nu\rangle$ need to vanish, which
could already be seen from Eq.~(\ref{Eq:B-mu-nu}) since for these
modes we have $z_\mu^{(1)}=0$. It can also easily be checked from
symmetry considerations that $|c_\perp\rangle$ is one of the modes as was
suggested earlier. The solution of the remaining two zeroth order
eigenfunctions is straightforward and leads to
\begin{align}
\label{Eq:viscomode}
|\psi_\perp^{(0)}\rangle &= |c_\perp\rangle,\\
\label{Eq:spm}
|\psi_{s_\pm}^{(0)}\rangle &=|s_\pm\rangle = 
  \frac{|s\rangle}{\sqrt{\langle s|s\rangle}} + \left(-\xi \pm
  \sqrt{1+\xi^2}\right) 
  \frac{|d\rangle}{\sqrt{\langle d|d\rangle}} ,
\end{align}
where the prefactors are determined by the requirements
\begin{align}
\langle s_\pm|s_\mp\rangle&=0,\\
\label{Eq:ortho-js}
\langle j_{s_\pm}| \frac{1}{\Omega} + \frac{1}{2}|j_{s_\mp}\rangle&= 0,
\end{align}
and we have introduced the short hand notation
\begin{equation}
\label{Eq:xi}
\xi \equiv\frac{\chi -{\mathcal D}}{2 Q},
\end{equation}
with
\begin{align}
\label{Eq:chi}
\chi &\equiv - \frac{\langle j_{s}| \frac{1}{\Omega} + \frac{1}{2} |j_{s}
\rangle}{\langle s|s\rangle},\\
\label{Eq:D}
{\cal D} &\equiv - \frac{\langle j_{d}| \frac{1}{\Omega} + \frac{1}{2} |j_{d}
\rangle}{\langle d|d\rangle},\\
\label{Eq:Q}
Q &\equiv - \frac{\langle j_{s}| \frac{1}{\Omega} + \frac{1}{2} |j_{d}
 \rangle}{\sqrt{\langle s|s\rangle \langle d|d\rangle}}.
\end{align}
As the form of the last three quantities suggests by comparison with
Eq.~(\ref{Eq:transport}), these quantities are related to transport
coefficients. In the low density limit and in the limit of only one
specie, one can identify $\chi$ as being the thermal diffusivity of
the GBL model. In the low density limit ${\cal D}$ will correspond to
the self-diffusion coefficient. Their interpretation as transport
coefficients is justified in Appendix \ref{Sec:appendixb}.
The remaining value $Q$ can not be interpreted as a transport
value, but rather is some measure of the coupling between
different modes.

Using Eq.~(\ref{Eq:transport}) we can evaluate the five transport
coefficients corresponding to the hydrodynamic modes. In the case of
the two soundmodes (\ref{Eq:soundmodes}) this leads to the
sounddamping $\Gamma$
\begin{equation}
z_\pm^{(2)} = \Gamma = - \frac{\langle j_\pm |\frac{1}{\Omega} + \frac{1}{2}
|j_\pm \rangle}{\langle \psi_\pm^{(0)} |\psi_\pm^{(0)}\rangle},
\end{equation}
while the perpendicular mode (\ref{Eq:viscomode}) gives rise to
the viscosity $\nu$
\begin{equation}
z_\perp^{(2)} = \nu = - \frac{\langle j_\perp| \frac{1}{\Omega} +
\frac{1}{2} 
|j_\perp \rangle}{\langle c_\perp|c_\perp\rangle}.
\end{equation}
By writing out the definitions of $|s_\pm\rangle$ and $|j_{s_\pm}
\rangle$ and using the introduced quantities (\ref{Eq:xi}) -
(\ref{Eq:Q}) where needed, we can rewrite the second order
eigenvalues of the two non-propagating modes $|s_\pm\rangle$ as 
\begin{equation}
 s^\circ_\pm =  - \frac{\langle j_{s_\pm}| \frac{1}{\Omega} +
\frac{1}{2} |j_{s_\pm} \rangle}{\langle s_\pm|s_\pm\rangle}=
\frac{1}{2} (\chi +{\cal D}) \pm Q \sqrt{1 + \frac{(\chi -
{\cal D})^2}{4 Q^2}}. 
\label{Eq:smodes}
\end{equation}
In the low density limit one finds that the two eigenmodes
 (\ref{Eq:spm}) up to a normalization factor will converge to 
$|s\rangle$ and $|d\rangle$ as defined in
 Eqns.~(\ref{Eq:Invariantbasis}) and
 (\ref{Eq:invariant-d}). Consequently one obtains $s^\circ_\pm
 \rightarrow \chi, {\cal D}$, which is an illustration of the
 decoupling of entropy and concentration fluctuations in this
 limit. In general, however, these two ``true'' 
transport coefficients do not seem to correspond with a conventional
transport coefficient, but rather they always appear in combination
with each other, and it is only in the appropriate limits that they
reduce to the thermal diffusivity and diffusion coefficient. This is
completely in agreement with the results known for the continuous
theory \cite{BoonYip}, where the coupling of fluctuations in
concentration and entropy results in the same effect.

In the present model, however, there is some ambiguity in the choice
made for the basic invariants. To arrive at this last formula it was
only necessary to assume that the invariants $|s\rangle$ and
$|d\rangle$, are mutually orthogonal, and span a 2-dimensional subspace
in the null-space which is perpendicular to $|c_\ell\rangle$,
$|c_\perp\rangle$, and $|p\rangle$. A good choice necessarily leads
to the proper modes in the fully known limit of one specie. Although this
puts some limitations on the possible choices it will not uniquely fix
the basis. The present choice for $|s\rangle$ in
(\ref{Eq:Invariantbasis}) is however the most natural extension of the
conventional formulation \cite{GBL93} and we will refer to $\chi$ as
being the generalized thermal diffusivity.

This problem is more easily identified in the diffusion-like mode
$|d\rangle$. In the present formulation (\ref{Eq:invariant-d}), it
contains in general a contribution proportional to $|p\rangle$, rather 
than being a weighted difference of the red and blue densities only as
found in the colored GBL model \cite{Blaak:2001PRE}. This already
suggests that this vector is not the proper generalization of the
self-diffusion mode. This is consistent with the continuous theory
\cite{BoonYip}, where the analogue of the $s_\pm^\circ$ also depend on
more than the thermal diffusivity and diffusion only. We will address
this subject again in a later section.

Similar to the case of the GBL model we can introduce $\gamma$, the
ratio of specific heats by
\begin{equation}
\gamma = 1 + \frac{\langle s|s\rangle}{\langle p|p\rangle} =
\frac{\langle p|p\rangle \langle \rho|\rho\rangle}{\langle p|\rho\rangle^2}
=\frac{c_s^2}{c_T^2},
\end{equation}
with $c_s^2=\langle p|p\rangle/\langle p|\rho\rangle$ the adiabatic
speed of sound, $c_T^2=\langle p|\rho\rangle/\langle \rho|\rho\rangle$
the isothermal speed of sound. Two additional useful definitions are
$\tau_{xy}=c_\ell c_\perp$ and $\tau_{xx}=\frac{1}{2}(c_\ell^2-
c_\perp^2)$. These allow us to rewrite the currents related to the
sound modes as $|j_\pm\rangle = |j_s\rangle \pm c_s |\tau_{xx}
\rangle$. Consequently this leads to 
\begin{equation}
\Gamma = - \frac{ \langle j_s |\frac{1}{\Omega} + \frac{1}{2} |j_s
\rangle + c_s^2 \langle \tau_{xx} |\frac{1}{\Omega} + \frac{1}{2}
|\tau_{xx} \rangle \pm 2 c_s^2 \langle j_s |\frac{1}{\Omega} +
\frac{1}{2} |\tau_{xx} \rangle}{\langle p|p\rangle + c_s^2 \langle
c_\ell|c_\ell\rangle \pm 2 c_s \langle p|c_\ell\rangle}.
\end{equation}
From  symmetry considerations it follows that $\langle p|c_\ell\rangle
= 0$ and $\langle j_s |\frac{1}{\Omega} + \frac{1}{2}
|\tau_{xx}\rangle = 0$, and since $\langle \tau_{xx} |\frac{1}{\Omega}
+ \frac{1}{2} |\tau_{xx} \rangle = \langle \tau_{xy} |\frac{1}{\Omega}
+ \frac{1}{2} |\tau_{xy} \rangle$ because of the isotropy of the
lattice, we obtain the following relation for the main transport
coefficients 
\begin{equation}
\Gamma=\frac{1}{2}\left(\nu+(\gamma-1)\chi\right).
\end{equation}

\begin{center}
\begin{figure}
\epsfig{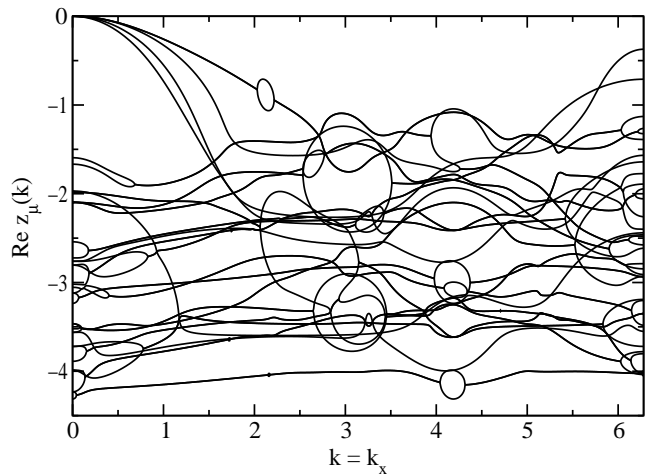}
\caption{\label{Fig:eigenvalue}
The real part of the full eigenvalue spectrum as a function of
the wavevector $|\bm{k}|$ in reciprocal lattice units. The system
parameters are $\rho=6.0$, $e=1.0$, and $P_r=0.75$. Note that there
are five hydrodynamic modes in the small wavevector limit. Two of
them, related to the soundmodes, coincide.} 
\end{figure}
\end{center}

An example of the full wave-vector dependent real eigenvalue spectrum
is shown in Fig.~\ref{Fig:eigenvalue}, which also confirms the
absence of spurious invariants. Of all eigenvalues only the five
related to the hydrodynamic modes go to zero in the limit  $\bm{ k}
\rightarrow 0$. It is with these modes that the binary-mixture
responds to deviations from thermal equilibrium 
\begin{align}
 z_\pm(\bm{ k})&= -\pm\imath c_s\bm{ k}-\Gamma \bm{ k}^2, \label{Eq:zpm}\\
 z_\perp(\bm{ k})&=-\nu \bm{ k}^2,\\
 z_{s_\pm}(\bm{ k})&=-s^\circ_{\pm} \bm{ k}^2. \label{Eq:zspm}
\end{align}
The first two eigenvalues describe sound propagation in the two
opposite directions parallel to $\bm{ k}$ with $c_s$ the adiabatic
sound speed, the third eigenvalue describes the shear mode, and the
last two eigenvalues represent purely diffusive, non-propagating
processes. In this hydrodynamic regime characterized by $k \lambda \ll
1$, where $\lambda$ is the mean free path length, one can exploit the
fact that the real component of the eigenvalues of the hydrodynamic
modes is much smaller than that of the kinetic modes. We will use this in 
the next section in order to obtain the Landau-Placzek formula 
(\ref{Eq:FullLP}) for the power spectrum.

For larger wavevectors (here roughly $0.5 \leq k \leq 1.5 $) the
relations (\ref{Eq:zpm})-(\ref{Eq:zspm}) start to deviate from the
true values. This is the generalized hydrodynamic regime ($k \lambda
\lesssim 1$), and the transport coefficients become $k$
dependent. The hydrodynamic modes are, however, still smaller than the
kinetic modes yielding a reasonable accurate  Landau-Placzek formula.
In the kinetic regime ($k \lambda \gtrsim 1$) the hydrodynamic modes
and kinetic modes become of the same order of magnitude. The
distinction between fast and slow modes can not be made and all modes
contribute to the power spectrum.

\begin{center}
\begin{figure}
\epsfig{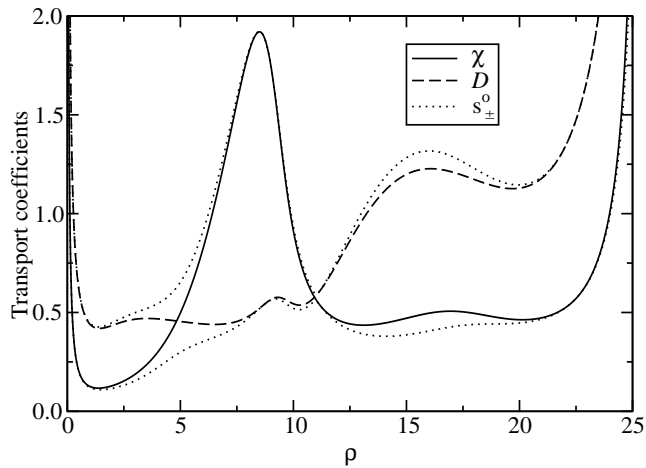}
\caption{\label{Fig:coupling} 
The ``true'' transport values $s^\circ_\pm$, and the
transport coefficients $\chi$, ${\cal D}$ as a function of the density
at $\theta=0.05$ and $P_r=0.75$. In the low density limit, and by the
duality of the model also in the high density limit, the two sets
converge.} 
\end{figure}
\end{center}

In Fig.~\ref{Fig:coupling} the diffusive transport properties are
shown at fixed reduced temperature
$\theta=\exp(-\frac{1}{2}\beta)$. The modes $s^\circ_\pm$ 
are the modes observed in the eigenvalue spectrum and are
combinations of $\chi$, ${\cal D}$, and $Q$. In the low density limit
these two modes converge to the thermal diffusivity and diffusion,
i.e. $s^\circ_+\rightarrow\chi$, $s^\circ_-\rightarrow{\cal
D}$, and is caused by the decoupling of the fluctuations in the
concentrations and entropy. In our model this means that the ratio
$Q/(\chi - {\cal D})$ vanishes. The value of $Q$, however, will 
in general remain small but finite due to the divergencies of the
transport coefficients in the low density limit of LGA. 
The same is observed in the high density limit, 
which is merely an illustration of the duality of the LGA-model if one
interchanges particles and holes. In both cases this is a direct
consequence of the fact that the fluctuations in the occupation numbers
become linear in the average occupation densities, thus leading to
equipartition.

As an additional remark we like to mention that the value of $Q$ can
be both positive and negative. Therefore the correct curves of
$s_\pm^\circ$ need not be continuous as a function of the density, but
can contain discontinuities located at points where $Q$ changes
sign. Consequently, depending on the system parameters, the role of
$s^\circ_+$ and $s^\circ_-$ is interchanged with respect to $\chi$ and
${\cal D}$. This is merely due to the choice in convention we have
used in Eq.~(\ref{Eq:spm}) and of no physical importance. 

For some intermediate values of the density one also finds
that the values of $s^\circ_\pm$ and $\chi, {\cal D}$ coincide. This
is, however, not caused by decoupling of the fluctuations. In these
cases there is no effective equipartition, but one obtains $Q=0$ as a
consequence of the cancellation of terms. Moreover, the location of
these points depends in a non-trivial way on the system parameters.

\begin{center}
\begin{figure}
\epsfig{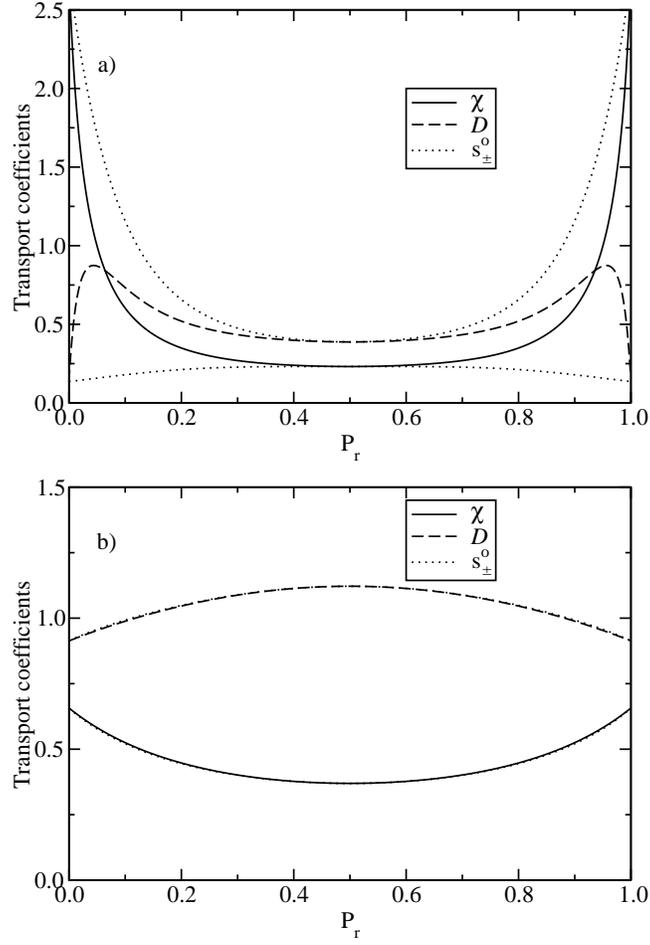}
\caption{\label{Fig:trans} 
The transport coefficients as a function of $P_r$ at (a)
$\rho=4$, $\theta=0.05$ and (b) $\rho=0.25$, $\theta=0.05$. For
low densities decoupling occurs for any composition, for higher
densities only in the limits $P_r=1$ and $P_r=1/2$.}
\end{figure}
\end{center}

The diffusive transport properties versus the relative concentration
of both species at fixed density and reduced temperature $\theta$ is
shown in Fig.~\ref{Fig:trans}. In the low density limit we have
$s_\pm^\circ\rightarrow \chi,{\cal D}$ for all relative concentrations. This
is not generally true as can be seen from the higher density
figure. It does, however, reveal that the decoupling also occurs in
general for the single specie limits $P_r\rightarrow0$ and
$P_r\rightarrow1$. This is not surprising, as in those limits the
model effectively reduces to a normal GBL model with only a single
diffusive mode related to the thermal diffusivity. This is actually
also the case in the high density limit of Fig.~\ref{Fig:coupling},
because there one has a situation in which the red lattice is almost 
completely filled and therefore an effective blue system that remains.

Finally, decoupling can also be observed in the case $P_r=P_b$. In fact
this is a rather special limit and could be analyzed completely in a
manner similar to the one used for the colored GBL model
\cite{Blaak:2001PRE}, because based on the symmetry of the problem one
can decompose the linearized Boltzmann operator in two type of
contributions, i.e. $|R\rangle+|B\rangle$ and $|R\rangle-|B\rangle$.

\begin{center}
\begin{figure}
\epsfig{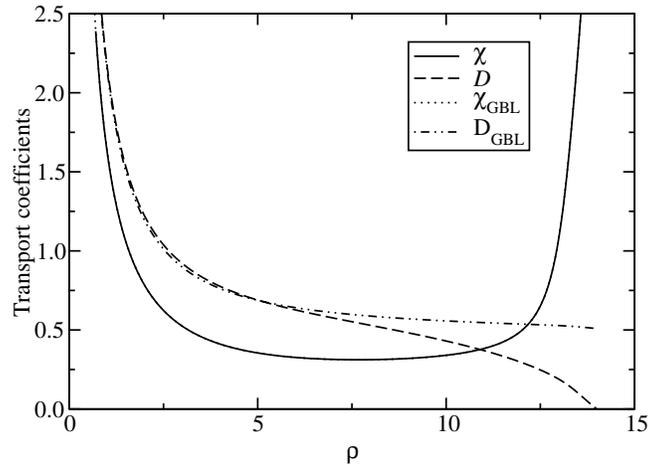}
\caption{\label{Fig:diff}
A comparison of the transport coefficients of the present model
in the limit $P_r\rightarrow 1$ with those of the GBL model for fixed
average energy density $e=1.0$. $\chi$ coincides with the thermal
diffusivity $\chi_{\text{GBL}}$ of the GBL model, ${\cal D}$ matches the  
self-diffusion $D_{\text{GBL}}$ only for the lower densities. The values
of $s^\circ_\pm$ are not shown but overlap completely with $\chi$ and {${\cal D}$}.}
\end{figure}
\end{center}

For $P_r=0$ the diffusion mode vanishes, since we then recover the
original GBL model, but for $P_r \rightarrow 0$ the diffusion mode
remains finite. This case is illustrated in Fig.~\ref{Fig:diff}. Here,
in the low density limit the diffusion mode becomes equal to the 
self-diffusion of the GBL model (and hence also equal to the diffusion
mode in the GBL color mixture), but for higher density they
deviate. A behavior that finds its origin in the fact that the color mixture
does not allow more than one particle in a single velocity channel and,
although the number of particles of the second specie gets very
small, they still have a large impact on the diffusion.

\section{Landau-Placzek theory}
\label{Sec:Landau-Placzek}

\subsection{Full spectrum}

In the hydrodynamic regime of small wavevectors ($k\rightarrow0$) and
the frequency 
$\omega$ being linear in $k$, the spectral density $S(\bm{
k},\omega)$ can be expanded in powers of $\imath k$. In this
long-wavelength, small frequency limit the hydrodynamic modes are well
separated from the kinetic modes, which can be neglected due to their
exponential decay. By keeping only terms up to $\bm{ O}(k^2)$ 
($\omega/k$ is kept constant for consistency), one obtains the
Landau-Placzek approximation.

It has been shown that the dynamic structure factor can be evaluated
by~\cite{GBL93} 
\begin{equation}
\frac{S(\bm{ k},\omega)}{S(\bm{ k})} = 2 \text{Re}
    \frac{\langle\rho|(e^{\imath\omega + 
    \imath\bm{ k}\cdot\bm{ c}}-1-\Omega)^{-1}+\frac{1}{2}|\rho\rangle}
  {\langle\rho|\rho\rangle}.
\end{equation}
The closely related spectral function $F(\bm{ k},\omega)$ can be
written as 
\begin{align}
\label{Eq:Fkw}
  F(\bm{ k},\omega)\equiv\langle\rho|(e^{\imath\omega+\imath\bm{ k}
 \cdot\bm{ c}} -1-\Omega)^{-1} + \frac{1}{2} |\rho\rangle = 
 2\sum_\mu\text{Re}{\cal N}_\mu {\cal D}_\mu,
\end{align}
where
\begin{align}
\label{Eq:Dmu}
{\cal D}_\mu &= \frac{1}{e^{\imath w - z_\mu}-1}+\frac{1}{2} \approx
\frac{z_\mu^{(2)} - \imath (w - z_\mu^{(1)}k) }{(z_\mu^{(2)}k^2)^2 +
(w - z_\mu^{(1)}k)^2},\\
\label{Eq:Nmu}
{\cal N}_\mu &= \langle \rho | \psi_\mu \rangle \langle \phi_\mu |
\rho \rangle. 
\end{align}
The coefficients ${\cal N}_\mu$ are evaluated in
Appendix~\ref{Sec:appendix} for small $\bm{ k}$ and yield  
\begin{align}
{\cal N}_\sigma & =  \frac{ \langle \rho|\rho\rangle}{2 \gamma} \left(1 +
\frac{\imath \sigma k}{c_s} \left[ \Gamma + (\gamma-1) \chi\right]
\right),\\
{\cal N}_{s_\pm} & =  \frac{(\gamma-1) \langle \rho | \rho
\rangle}{2 \gamma} \left( 1 \pm \frac{\chi-{\cal
D}}{s^\circ_+-s^\circ_-} \right).
\end{align}
From symmetry considerations one can conclude that the shear mode will
not contribute to the spectrum and one finds ${\cal
N}_{\perp}=0$. Combining these results with the expressions
(\ref{Eq:Fkw}) and (\ref{Eq:Dmu}), we finally obtain the
Landau-Placzek formula, describing the power-spectrum in the
hydrodynamical domain
\begin{eqnarray}
& \frac{S(\bm{ k},\omega)}{S(\bm{ k})} = \frac{\gamma-1}{\gamma}
\left[ \left(1+\frac{\chi-{\cal D}}{s^\circ_+-s^\circ_-}\right)
\frac{s^\circ_+ k^2}{\omega^2+(s^\circ_+ k^2)^2}+
\left(1-\frac{\chi-{\cal D}}{s^\circ_+-s^\circ_-}\right)
\frac{s^\circ_- k^2}{\omega^2+(s^\circ_- k^2)^2} \right] \nonumber \\
& +\frac{1}{\gamma}\left[\frac{\Gamma k^2}{(\omega+c_s k)^2+(\Gamma
k^2)^2} +\frac{\Gamma k^2}{(\omega-c_s k)^2+(\Gamma k^2)^2}\right]
\nonumber \\
&+\frac{1}{\gamma}\left[\Gamma+(\gamma-1)\chi\right]\frac{k}{c_s}
\left[\frac{\omega+c_s k}{(\omega+c_s k)^2+(\Gamma k^2)^2}-
\frac{\omega-c_s k}{(\omega-c_s k)^2+(\Gamma k^2)^2}\right].
\label{Eq:FullLP}
\end{eqnarray}
The spectrum contains an unshifted central peak that is formed by two
Lorentzians due to the two processes related to the non-propagating
modes $s_\pm^\circ$. The two
propagating modes lead to the presence of the two shifted Brillouin
lines. Their width at half-height can be used as a measurement of
$\Gamma \bm{ k}^2$, the position of the peaks can be used a
measurement of $\pm c_s \bm{ k}$. The last two terms in
Eq.~(\ref{Eq:FullLP}) give an asymmetric correction to the
Brillouin peaks and induce a slight pulling of the peaks toward the
central peak. 

The symmetry of the different contributions is such that the ratio of
the integrated contributions of the central peak and the Brillouin
components is constant and given by
\begin{equation}
\frac{\sum_\sigma \int d \omega Re({\cal N}_{s_\sigma} {\cal
D}_{s_\sigma})}{\sum_\sigma \int d \omega Re({\cal N}_\sigma {\cal
D}_\sigma)} = \gamma - 1.
\end{equation}

\begin{center}
\begin{figure}
\epsfig{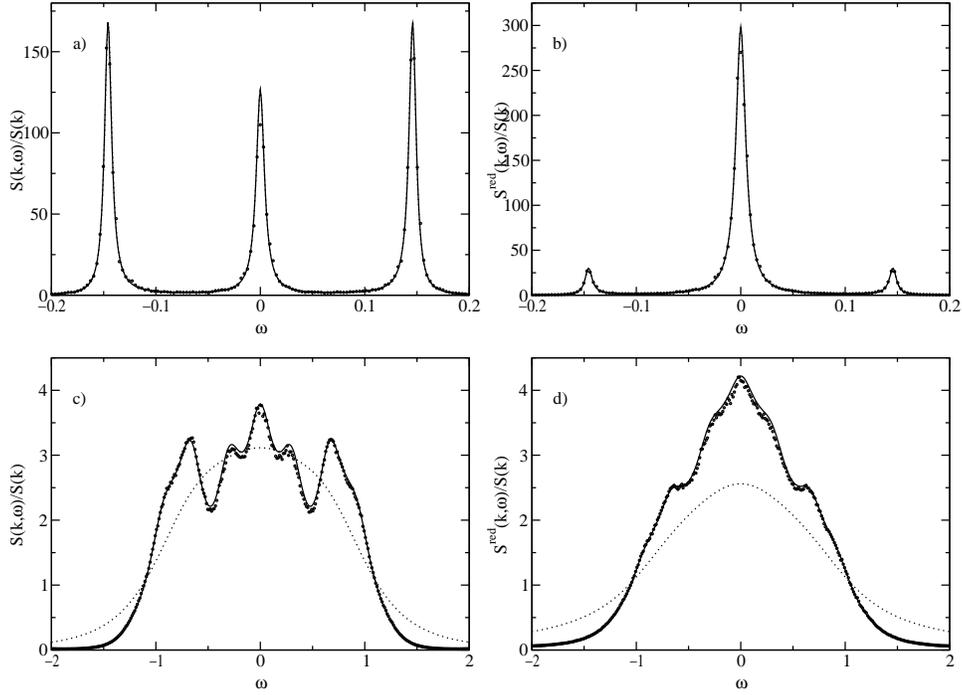}
\caption{\label{Fig:spectra}
Boltzmann spectra and Landau-Placzek approximations for the full
(a,c) and for the red spectrum (b,d). The system parameters are
$\rho=10$, $P_r=0.25$, $e=0.75$, $k_x=10\times2\pi/512$ (a,b) and 
$\rho=1$, $P_r=0.35$, $e=1.0$, $k_x=52\times2\pi/512$ (c,d). The solid
and dotted curves are the Boltzmann results and Landau-Placzek
approximations respectively. Simulations results are indicated by the
points. The wavevector $\bm{ k}$ is given in reciprocal lattice
units, the frequency $\omega$  in reciprocal time ($2\pi/T$ with $T$
the total number of time steps), and the spectral functions in reciprocal
$\omega$ units. } 
\end{figure}
\end{center}

In Fig.~\ref{Fig:spectra} the Landau-Placzek formula is compared with
the full Boltzmann spectrum for two different wavevectors. For the
smallest wavevector inside the hydrodynamic regime they coincide,
while for the larger they differ considerably and indicates that this
wavevector is outside the hydrodynamic, and in fact inside the
kinetic regime.

\begin{center}
\begin{figure}
\epsfig{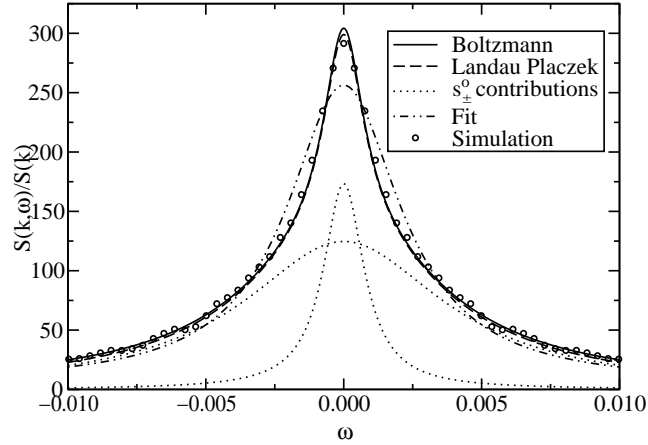}
\caption{\label{Fig:central} 
The central peak of the full spectrum at $\rho=6.5$, $P_r=0.05$,
$\theta=0.05$, and $k_x=4\times2\pi/512$. The Boltzmann spectrum and
Landau Placzek formula overlap almost completely. The two isolated
contributions of the central peaks in the later are indicated as
well. The fit is made on the simulation results (points) with a single
central peak only. Note that the two sound peaks fall outside
the interval shown here.}
\end{figure}
\end{center}

In general the contributions forming the central peak cannot easily be
separated. Even in the case they differ sufficiently in order to
fit the central peak with two Lorentzians, one only obtains
information on the values of $s^\circ_\pm$, which is not enough to
determine the more interesting values of the thermal diffusivity and
diffusion. An example of this is shown in Fig.~\ref{Fig:central} where
we decomposed the central peak in the two Lorentzians using the
theoretical expressions. For comparison
we included the least square fit according to Eq.(\ref{Eq:FullLP}) to
the full range of the spectrum, including the Brillouin peaks, where
we only used a single Lorentz for the central peak. 

For this reason it is important to consider the limits where
decoupling occurs, since in those cases the the relevant transport
values can be obtained from the spectra. Here we have however an
interesting difference with respect to light scattering
experiments~\cite{BoonYip}. Whereas in that case the sensitivity of
the dielectric fluctuations with respect to density is much larger
than with respect to temperature, the amplitude factors will differ
orders of magnitude. This results in the observation that the
diffusion component will usually dominate the spectrum. For LGA this 
does not apply, and by using the decoupling limit $s^\circ_\pm
\rightarrow \chi, {\cal D}$ on the first term off the Landau-Placzek
formula (\ref{Eq:FullLP}), one immediately obtains that the central
component of the spectrum is given by a single peak characterized by
the thermal diffusivity
\begin{equation}
\frac{S^{\text{cen}}(\bm{ k},\omega)}{S(\bm{ k})} =
\frac{\gamma-1}{\gamma}
\frac{2 \chi k^2}{\omega^2+(\chi k^2)^2}.
\end{equation}
One could in principle use this also in some of the intermediate cases
seen in Fig.~\ref{Fig:coupling}. The location of these points in terms
of density and relative fractions, however, depends in a non-trivial
way on the system parameters. Moreover, the identification with the
thermal diffusivity and diffusion can only be made if one interpret
$\chi$ and ${\cal D}$ as generalizations of these quantities. 

\subsection{Red spectrum}
The red dynamic structure factor can be evaluated
by~\cite{Hanon,Blaak:2001PRE} 
\begin{equation}
\frac{S^{\text{red}}(\bm{ k},\omega)}{S(\bm{ k})}=2\text{Re} \frac{\langle
  R|(e^{\imath\omega+ 
  \imath\bm{ k}\cdot\bm{ c}}-1-\Omega)^{-1}+\frac{1}{2}|R\rangle}
  {\langle R| R\rangle}.
\end{equation}
It is possible to follow the same route for the red dynamic structure
factor as for the full spectrum. However, since $\langle
\rho|d\rangle=0$, but $\langle R|d\rangle\not=0$, it is more
convenient to use a different set of basic invariants
\begin{equation}
 |a_n \rangle =\{ |c_x\rangle, |c_y\rangle,
 |p\rangle = |\frac{1}{2} c^2\rangle, |s_r\rangle = |p\rangle - 
 \frac{\langle p|p\rangle}{\langle p|R\rangle} |R\rangle,
 |d_r\rangle\},
\end{equation}
where $|d_r\rangle$ is constructed to be perpendicular to the other
conserved quantities and defined by 
\begin{equation}
|d_r\rangle = |p\rangle-\left(\frac{\langle p|R\rangle}{\langle
R|R\rangle} |R\rangle 
 -\frac{\langle p|R\rangle^2-\langle p|p\rangle\langle R|R\rangle}
  {\langle R|R\rangle\langle p|B\rangle}|B\rangle\right).
\end{equation}
Note that by construction it is perpendicular to the red density,
i.e. $\langle R|d\rangle = 0$. Analogous to the case of the normal density we
introduce some transport-like coefficients related to this basis
\begin{align}
 \chi_{\text{r}}&=-\frac{\langle s_r|\frac{1}{\Omega}+\frac{1}{2}|s_r\rangle}
    {\langle s_r|s_r\rangle},\\
 {\cal D}_{\text{r}}&=-\frac{\langle
    d_r|\frac{1}{\Omega}+\frac{1}{2}|d_r\rangle} 
    {\langle d_r|d_r\rangle},\\
 \gamma_{\text{r}}&=1+\frac{\langle s_r|s_r\rangle}{\langle p|p\rangle}.
\end{align}
Obviously, the sound damping $\Gamma$, the modes $s^\circ_\pm$, and
the speed of sound $c_s$ are all unchanged, since these are true
transport values and independent of a chosen set of basis
functions. Also in this case we can derive a Landau-Placzek formula,
describing the red power-spectrum in the hydrodynamical domain
\begin{eqnarray}
& \frac{S^{\text{red}}(\bm{ k},\omega)}{S(\bm{ k})} =
 \frac{\gamma_{\text{r}}-1}{\gamma_{\text{r}}}\left[
 \left(1+\frac{\chi_{\text{r}}-{\cal
 D}_{\text{r}}}{s^\circ_+-s^\circ_-}\right) 
 \frac{s^\circ_+ k^2}{\omega^2+(s^\circ_+ k^2)^2}+
 \left(1-\frac{\chi_{\text{r}}-{\cal
 D}_{\text{r}}}{s^\circ_+-s^\circ_-}\right) 
 \frac{s^\circ_- k^2}{\omega^2+(s^\circ_- k^2)^2}\right] \nonumber \\
& +\frac{1}{\gamma_{\text{r}}}\left[\frac{\Gamma k^2}{(\omega+c_s
 k)^2+(\Gamma k^2)^2} 
 +\frac{\Gamma k^2}{(\omega-c_s k)^2+(\Gamma k^2)^2}\right] \nonumber \\
& +\frac{1}{\gamma_{\text{r}}}\left[(\nu-\Gamma)+
 2(\gamma_{\text{r}}-1)\chi_{\text{r}}    
 \right]\frac{k}{c_s}\left[\frac{\omega+c_s k}{(\omega+c_s k)^2+
 (\Gamma k^2)^2}- 
  \frac{\omega-c_s k}{(\omega-c_s k)^2+(\Gamma k^2)^2}\right].
\end{eqnarray}
In principle the quantities $\chi_{\text{r}}$, ${\cal
D}_{\text{r}}$, and $\gamma_{\text{r}}$ can be expressed in the
normal transport values. However, these relations lead to more complex
expressions and are for that reason omitted here. In
Fig.~\ref{Fig:spectra} this formula is compared with the full Boltzmann
red spectrum for a wave vector in the hydrodynamic regime, yielding a
satisfactory approximation, and one in the kinetic regime with a large
discrepancy. 

\subsection{Diffusion spectrum}
\label{Sec:diffusion}
Finally we like to consider the spectra of the diffusive processes
only. Due to the nature of LGA, however, it is not obvious 
how to define diffusion properly. On the one hand we have the
requirement that a diffusive process is not propagating, on the other
hand one expects the diffusion only to be related to the densities of
both components in the mixture. As we will show in this section, it
turns out that these two views are not completely compatible.

In the colored GBL model \cite{Blaak:2001PRE} and in the
continuous case \cite{McLennan} a diffusion spectrum 
 can be obtained by considering fluctuations in the
normalized difference of the red and blue density
\begin{equation}
\rho_{\text{diff}} = \frac{\rho_{\text{red}}}{P_r} -
\frac{\rho_{\text{blue}}}{P_b}.
\end{equation}
The proper translation in terms of invariants is given by
\begin{equation}
\label{Eq:diff-0}
|\text{diff}\rangle = \frac{|R\rangle}{P_r} - \frac{|B\rangle}{P_b}.
\end{equation}
However, a spectrum based on this vector, does not lead to purely
diffusive peaks only, but also includes parts of the propagating
modes. This can easily be checked since the vector is in general not
orthogonal to the two soundmodes due to the absence of equipartition.  

A simple solution is to subtract the propagating part by
adding the appropriate term proportional to $|p\rangle$ 
\begin{equation}
\label{Eq:diff-1}
|\text{diff}\rangle = \frac{|R\rangle}{P_r} - \frac{|B\rangle}{P_b} -
\left( \frac{\langle p|R\rangle}{P_r} - \frac{\langle
p|B\rangle}{P_b}\right) \frac{|p\rangle}{\langle p|p\rangle}.
\end{equation}
In a binary athermal mixture \cite{Ernst:1990JSP}, however, it was suggested
to use  
\begin{equation}
|\text{diff}\rangle = \frac{|R\rangle}{\langle R|R \rangle} -
\frac{|B\rangle}{\langle B|B\rangle }.
\label{Eq:diffErnst}
\end{equation}
Unfortunately this choice also leads to a propagating mode in the
thermal case. One could again subtract the propagating part, but the
result would be different from (\ref{Eq:diff-1}). The origin of this
problem is that we have two different diffusive modes and any linear
combination would lead to a diffusive spectrum. However, due to
the lack of equipartition the different choices of fluctuations one
wishes to consider do not coincide, not even after the propagating part
is eliminated. 

Naively one would expect a combination of the red and blue component
only and in addition it should be perpendicular to the propagating
modes. This leads to the following generalization of the athermal
result (\ref{Eq:diffErnst}), which in the case of an athermal model would
coincide  
\begin{equation}
\label{Eq:diff-2}
|\text{diff}\rangle = \frac{|R\rangle}{\langle p|R\rangle} -
 \frac{|B\rangle}{\langle p|B\rangle}.
\end{equation}
Obviously there is some freedom here in order to choose the
generalization of the diffusion. The natural extension would be one
that that satisfies the appropriate limits. However, in the low density
limit and in the limit $P_r \rightarrow 1$, they all converge to the same
value equal to the one found for the single component GBL model. The
special case of an equal density for red and blue particles is not
helpful either. This limit can be completely analyzed in a manner
analogous to what is done for colored GBL model
\cite{Blaak:2001PRE}. For reasons of symmetry this will cause 
none of the definitions to have a propagating character and the
resulting diffusions would coincide.

\begin{center}
\begin{figure}
\epsfig{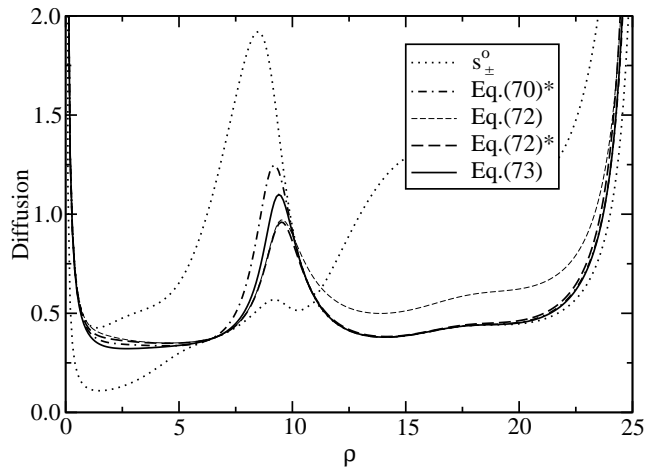}
\caption{\label{Fig:diffs} 
A comparison of the different diff's as a function of
the density at $\theta=0.05$ and $P_r=0.75$. The curves labeled by a *
are corrected by subtracting the propagating part as is described in
the text. The diffusion according to Eq.~(\ref{Eq:diff-0}) is not
shown explicitly but overlaps on this scale with its corrected
counterpart. }
\end{figure}
\end{center}

In Appendix \ref{Sec:appendixb} we derive the proper definition
(\ref{Eq:order2}) for the transport values related to these
quantities. Since none of them are proportional to a single eigenmode
the usual formulation (\ref{Eq:transport}) is no longer valid. The
results can be found in Fig.~\ref{Fig:diffs}, where the diffusions
obtained for the various diff's are shown as function of the density. 
In general the five different diff's described above all
lead to different values for the corresponding transport value,
although depending on the choice of system parameters this difference
might be marginal. Also compare with Fig.~\ref{Fig:coupling} to see
the difference with respect to ${\cal D}$.

Although the choice (\ref{Eq:diff-2}) is the most natural
generalization, the diffusion is only properly identified in the
limiting cases.  In general some arbitrariness remains in a binary
thermal lattice gas. 

\begin{center}
\begin{figure}
\epsfig{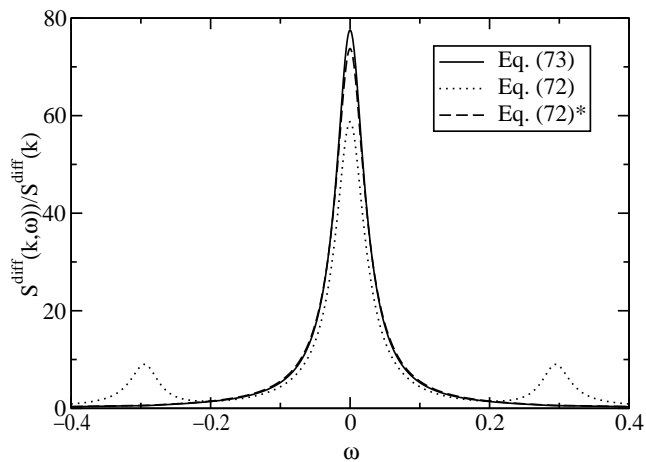}
\caption{\label{Fig:diffspectrum} 
The Boltzmann diffusion spectrum based on
Eqns.~(\ref{Eq:diffErnst}) and (\ref{Eq:diff-2}). The curve labeled by
a * is corrected by subtracting the propagating part as is described
in the text. The system parameters are $P_r=0.15$, $\rho=10.0$, and
$\theta=0.05$. The wavevector $\bm{ k}$ is given in reciprocal
lattice units, the frequency $\omega$ in reciprocal time.}
\end{figure}
\end{center}

The corresponding diffusion spectrum is obtained by
\begin{equation}
\frac{S(\bm{ k},\omega)}{S(\bm{ k})}=2\text{Re} \frac{\langle
  \text{diff}|(e^{\imath 
  \omega+ \imath\bm{ k}\cdot\bm{ c}}-1-\Omega)^{-1} + \frac{1}{2}|
  \text{diff} \rangle}{\langle \text{diff} | \text{diff} \rangle}.
\end{equation}
and shown in Fig.~\ref{Fig:diffspectrum} for two diff's:
the $|\text{diff}\rangle$ defined in Eq.~(\ref{Eq:diffErnst}) based on
an athermal model and the correct adjustment for the thermal
model Eq.~(\ref{Eq:diff-2}). For comparison we also included the
curve for which the propagating part in  Eq.~(\ref{Eq:diffErnst}) is
subtracted as was described above. The first diffusion spectrum has a
Brillouin-like pair of peaks which is a manifestation of the
propagating part of $|\text{diff}\rangle$ according to
(\ref{Eq:diffErnst}), while the other two are different superpositions
of two Lorentzians characterized by ${s_\pm}^{(0)}$.

\section{Simulation results}
\label{Sec:simulation}

We have verified our results with simulation results. Some notes on
the implementation are in order. The model we consider is a true
$38$-bits model, i.e. the collision operator can act on $2^{38}$
different 
states. In a previous article we reported the simulation results of a
colored GBL model \cite{Blaak:2001PRE}. The number of states in that
model was $3^{19}$, since no red and blue particle could exist with the
same velocity. The simulations could, however, be performed by
realizing that the collisions could be separated in a normal GBL
collision followed by a redistribution of the colors over the
different particles present. These two separated processes can easily
be performed by any computer.

In the present model this approach can not be used. One could of
course make use of the symmetries present in the model. The hexagonal
lattice leads to six rotations and two reflections. In addition we can
make use of the interchange of red and blue particles, and the
interchange of particles and holes. At best this would lead to a
reduction with a factor of 48 on the total number of different
states, but in practice this factor is less because a large fraction
of the states is invariant with respect to some of these symmetries. This is
however still too large in order to be applied in any but the largest
supercomputers available at present. An alternative, but rather
inefficient scheme, would be to store only part of the collision table
and generate other collisions on the fly. 

Binary mixtures, however, do allow for a more convenient solution,
which is less efficient than storing the complete collision table, but
still has  a relative good performance (about a factor 2-5 slower than
the colored GBL model).
Rather than storing a collision
table based on the states, we make one based on the different
classes. In the GBL model there are $29926$ different classes ${\cal
C}=(M,\bm{ P},E)$, characterized by the total mass, momentum, and
energy. Since the the binary mixture is the combination of two GBL
models there are $29926^2$ different combinations $({\cal
C}_R,{\cal C}_B)$ of a red and blue class which form a total of
$4478568$ classes $\Gamma=(M_r,M_b,\bm{ P},E)$. If we combine this
with the 48 symmetry operations we get a working algorithm that
already can be used on a computer with 256 MB of memory.

An arbitrary input state is now analyzed in order to determine to
which class $\Gamma$ it belongs. From its characterization one can
easily find which transformation is needed in order put it in a form
with $M_R \leq M_B$, $M_R+M_B\leq 19$, and $0 \leq P_y
\leq P_x/\sqrt{3}$. The last inequality confines the total momentum to
an angle of $\pi/6$. For each $\Gamma$ in this limited set of classes
only, all the classes ${\cal C}_R$, that with the appropriate class
${\cal C}_B$ can give rise to it, have been stored, including the
number of states in each $({\cal C}_R,{\cal C}_B)$ combination. From
this we generate with the proper weight an ``outgoing'' class
combination to which the inverse transformation is
applied. Finally we only need to determine a random state in ${\cal
C}_R$ and ${\cal C}_B$, which is just a GBL-like process.

Simulation results along with the Boltzmann approximation and the
Landau-Placzek formula are shown in Fig.~\ref{Fig:spectra} for both
the normal and red spectrum. In the Figs.~\ref{Fig:spectra}a and
\ref{Fig:spectra}b the spectra  at low $\bm{ k}$ value completely 
overlap and the Landau-Placzek formula describes the power spectrum
very well. If we increase the wavevector $\bm{ k}$, and/or
lower the density, the hydrodynamical regime is left and deviations
start to appear. In that generalized regime, the transport properties
are $\bm{ k}$-dependent. Figs.~\ref{Fig:spectra}c and
\ref{Fig:spectra}d show a spectrum even   
further away from the hydrodynamic regime, in the kinetic regime. Several 
kinetic modes invade the spectrum, and a parameterization into
Lorentzians has lost all physical meaning. The full spectrum based on
the Boltzmann approximation, however, still leads to a very good
description, supporting the molecular chaos assumption.

\section{Discussion}
\label{Sec:discussion}

We have constructed a thermal binary lattice gas mixture. The model is
characterized by cross-effects between energy transport and
diffusion. The Landau-Placzek formula derived in the the low
wavevector, low frequency domain gives an excellent description. For
larger wavevectors it will fail, but the spectra can still accurately
be described by the Boltzmann approximation.

The Landau-Placzek formula for a regular binary thermal mixture is
quite similar in structure as the one for the continuous case. The main
feature is a central peak formed by two Lorentzians due to the coupled
entropy-concentration fluctuations. In the limits of low density,
single specie, their dual interpretations, and equal red and blue
density it is possible for the central peak of the spectrum to be
decomposed into two Lorentzians with linewidths given by $\chi\bm{
k}^2$ and ${\cal D}\bm{ k}^2$, otherwise the linewidths depend on
both these values.  

In the low density limit the true modes $s^\circ_\pm$ converge to
$\chi$ and ${\cal D}$. Moreover, they can be identified with the thermal
diffusivity and mass-diffusion coefficient of the GBL model 
and coincide with the low density limit of the continuous binary
mixture. In contrast with continuous theory, however, it is the
thermal diffusivity that dominates the central part of the spectrum,
rather than the diffusion. The interpretation of $\chi$ as the thermal
diffusivity can be extended to the general situation. For ${\cal D}$
this not true, since the related spectrum will contain Brillouin
peaks, which is a not a purely diffusive property. Although it is
possible to correct for this, there remains some ambiguity 
in the choice for the generalized diffusion coefficient.

The analysis and results presented here are in general true for binary
thermal lattice gasses and not restricted to this particular model
only. It also reduces automatically to the proper formulation for an
athermal model, in which case some of the ambiguities are removed.

\section*{Acknowledgments}
We would like to thank H.~Bussemaker and D.~Frenkel for helpful discussions.
R.B. acknowledges the financial support of the EU through the
Marie Curie Individual Fellowship Program (contract no.~HPMF-CT-1999-00100).

\appendix
\section{}
\label{Sec:appendix}

To calculate terms of the type ${\cal N}_\mu$ given in
Eq.~(\ref{Eq:Nmu}) we express the left eigenvectors in terms of right
eigenvectors, and expand ${\cal N}_\mu$ to linear order in $k$
\begin{eqnarray}
{\cal N}_\mu = & \frac{\langle \rho | \psi_\mu^{(0)}
\rangle^2}{\langle \psi_\mu^{(0)} 
| \psi_\mu^{(0)} \rangle} \left( 1 + 2 \imath k \left[ \frac{\langle
\rho | \psi_\mu^{(1)} \rangle}{\langle \rho | \psi_\mu^{(0)} \rangle}
- \frac{\langle \psi_\mu^{(0)} | \psi_\mu^{(1)} \rangle}{\langle
\psi_\mu^{(0)} | \psi_\mu^{(0)} \rangle} \right] + \right. \nonumber
\\ & \left. \imath 
k \left[ \frac{\langle \rho | c_\ell \psi_\mu^{(0)} \rangle}{\langle \rho
\psi_\mu^{(0)} \rangle} - \frac{\langle \psi_\mu^{(0)} | c_\ell
\psi_\mu^{(0)} \rangle}{\langle \psi_\mu^{(0)} | \psi_\mu^{(0)}
\rangle} \right] + {\cal O}(k^2) \right).
\label{Eq:N-mu}
\end{eqnarray}
As can be seen from the second term on the righthand side, partial
knowledge of the $|\psi_\mu^{(1)}\rangle$ is required. From the formal
solution (\ref{Eq:solution-one}) of the first order equation and the
orthogonality relation (\ref{Eq:psi0-j-m}) it follows that we only
need to evaluate the coefficients $B_{\mu \nu}$. From the
normalization (\ref{Eq:normalisation}) we already know that
$B_{\mu\mu}=0$. In addition we like to mention that from symmetry
considerations one can conclude that the mode related to the viscosity
will not contribute. Hence the values of $B_{\mu\perp}$ and
$B_{\perp\mu}$ need not be determined. 

In order to facilitate the calculations we first list some simple
relations
\begin{equation}
\langle \psi_\sigma^{(0)}|\psi_\sigma^{(0)}\rangle = 2 \langle
p|p\rangle,
\end{equation}
\begin{equation}
\langle \rho|\psi_\sigma^{(0)}\rangle =\langle \rho|p\rangle,
\end{equation}
\begin{equation}
\langle \rho|s_\pm \rangle = \frac{(1-\gamma)}{\sqrt{\langle
s|s\rangle}} \langle \rho|p\rangle,
\end{equation}
\begin{equation}
\langle s_\pm|s_\pm\rangle = 2 \sqrt{1+\xi^2}
\left(\sqrt{1+\xi^2} \mp \xi \right). 
\end{equation}
Most coefficients can be evaluated from the relation
(\ref{Eq:B-mu-nu}). In the case of the soundmodes this leads to 
\begin{equation}
B_{\sigma,-\sigma} = - \frac{\langle j_{-\sigma} | \frac{1}{\Omega} +
\frac{1}{2}| j_\sigma\rangle}{(z_\sigma^{(1)} - z_{-\sigma}^{(1)})
\langle \psi_{-\sigma}^{(0)}|\psi_{-\sigma}^{(0)}\rangle}
\end{equation}
and with the use of $|j_\pm\rangle = |j_s\rangle \pm c_s
|\tau_{xx}\rangle$ this can be rewritten in terms of the basic
transport coefficients
\begin{equation}
B_{\sigma,-\sigma} = \sigma \frac{\nu - (\gamma-1)\chi}{4 c_s} =
\frac{\sigma (\nu - \Gamma)}{2 c_s}. 
\end{equation}
The next coefficient we need to evaluate is 
\begin{equation}
B_{s_\pm,\sigma} = - \frac{\langle j_\sigma | \frac{1}{\Omega} +
\frac{1}{2}| j_{s_\pm}\rangle}{(z_{s_\pm}^{(1)} - z_\sigma^{(1)})
\langle \psi_\sigma^{(0)}|\psi_\sigma^{(0)}\rangle}, 
\end{equation}
and here we can use the same relation to rewrite the current of the
soundmodes. For the other current the relation (\ref{Eq:spm}) can be
used. Realizing that the terms containing $|\tau_{xx}\rangle$ will
vanish, the remaining terms can be manipulated to yield
\begin{equation}
B_{s_\pm,\sigma} = (\gamma - 1) \frac{ \sigma s^\circ_\pm }{2 c_s
\sqrt{\langle s|s\rangle}}. 
\end{equation}
From Eq.~(\ref{Eq:B-mu-nu}) we can now also see that
\begin{equation}
B_{\sigma,s_\pm} = - \frac{\langle \psi_\sigma^{(0)} |\psi_\sigma^{(0)}
\rangle}{\langle s_\pm|s_\pm\rangle} B_{s_\pm,\sigma} = 
-\left(1 \pm \frac{\xi}{\sqrt{1 + \xi^2}}\right)\frac{\sigma
s^\circ_\pm \sqrt{\langle s|s\rangle}}{2 c_s}. 
\end{equation}

The last two coefficients which we will need are
$B_{s_\pm,s_\mp}$. However, since $z_{s_\pm}^{(1)}=0$ they can not be 
obtained from (\ref{Eq:B-mu-nu}). In order for them to be determined
we need to make use of the third order equation of the eigenvalue
problem (\ref{Eq:eigenproblem-right})
\begin{eqnarray}
\Omega |\psi_\mu^{(3)}\rangle & = & (c_\ell + z_\mu^{(1)})
|\psi_\mu^{(2)}\rangle + \left[ z_\mu^{(2)} + \frac{1}{2} (c_\ell +
z_\mu^{(1)} )^2 \right] |\psi_\mu^{(1)}\rangle + \nonumber \\ 
&& \left[z_\mu^{(3)} +
z_\mu^{(2)}(c_\ell + z_\mu^{(1)}) + \frac{1}{6} (c_\ell +
z_\mu^{(1)})^3  \right] |\psi_\mu^{(0)}\rangle.
\end{eqnarray}
As we only want to determine the values of $B_{s_\pm,s_\mp}$, we do
not attempt to solve the complete third order equation but restrict
ourselves to the two equations from which they can be obtained.
Note that we have $z_{s_\pm}^{(1)}=0$ and hence $|j_{s_\pm}\rangle =
|c_\ell s_\pm\rangle$. 

Substituting these results in the third order equation, replacing
$\mu$ by $s_\pm$, and multiplying on the left with the appropriated
term $\langle s_\mp|$ we find
\begin{equation}
\label{Eq:order-three}
0 = \langle s_\mp |
c_\ell | \psi_{s_\pm}^{(2)}\rangle + \langle s_\mp | s^\circ_\pm +
\frac{1}{2} c_\ell^2 | \psi_{s_\pm}^{(1)}\rangle + \langle s_\mp |
z_{s_\pm}^{(3)} + s^\circ_\pm c_\ell + \frac{1}{6} c_\ell^3 |
s_\pm \rangle.
\end{equation}
The last term will disappear due to the odd power of $c_\ell$ and  
because $\langle s_\mp | s_\pm \rangle = 0$. In order to proceed we
not only need the solution of the first order equation
(\ref{Eq:solution-one}) but also solution of the second order
equation. Fortunately the later does not have to be computed
completely but the formal solution will suffice
\begin{eqnarray}
& |\psi_\mu^{(2)}\rangle = \frac{1}{\Omega} (c_\ell + z_\mu^{(1)})|
 \psi_\mu^{(1)}\rangle + \frac{1}{\Omega}\left( z_\mu^{(2)} +
 \frac{1}{2}(c_\ell + z_\mu^{(1)})^2 \right) | \psi_\mu^{(0)}\rangle +
\nonumber \\ & \sum_\nu C_{\mu \nu} |\psi_\nu^{(0)}\rangle,
\end{eqnarray}
where the $C_{\mu \nu}$ are some unknown coefficients. Substitution of
both solutions in Eq.~(\ref{Eq:order-three}) gives
\begin{eqnarray}
& 0 = \langle s_\mp | c_\ell \frac{1}{\Omega}  c_\ell|
\psi_{s_\pm}^{(1)}\rangle + \langle s_\mp | c_\ell \frac{1}{\Omega}
\left( s^\circ_\pm + \frac{1}{2} c_\ell^2 \right) | s_\pm \rangle +
\sum_\nu C_{s_\pm \nu} \langle j_{s_\mp} | \psi_\nu^{(0)}\rangle +
\nonumber \\ &
\langle s_\mp | \left( s^\circ_\pm + \frac{1}{2} c_\ell^2 \right)
\frac{1}{\Omega} | c_\ell s_\pm\rangle + \sum_\nu B_{s_\pm \nu}
\langle
s_\mp | \left( s^\circ_\pm + \frac{1}{2} c_\ell^2 \right)
|\psi_\nu^{(0)}\rangle.
\end{eqnarray}
The second and fourth term on the right are zero because of the odd
power in $c_\ell$, the third term is zero because of the orthogonality
relations (\ref{Eq:psi0-j}). Another substitution of the first order
solution leads to 
\begin{eqnarray}
0 = & \langle s_\mp | c_\ell \frac{1}{\Omega}  c_\ell
\frac{1}{\Omega}  c_\ell | s_\pm \rangle + \sum_\nu B_{s_\pm \nu}
\langle s_\mp | c_\ell \frac{1}{\Omega} c_\ell | \psi_\nu^{(0)}
\rangle + \nonumber \\ & \sum_\nu B_{s_\pm \nu} \langle
s_\mp | \left( s^\circ_\pm + \frac{1}{2} c_\ell^2 \right)
|\psi_\nu^{(0)}\rangle.
\end{eqnarray}
The first term is again zero because of the odd power in
$c_\ell$. Writing out the sum, using the orthogonality relations, and
realizing that the shear mode dependent term cancels due to symmetries
this leads to
\begin{eqnarray}
&& B_{s_\pm,s_\mp} s^\circ_\pm \langle s_\mp | s_\mp \rangle +
B_{s_\pm,s_\mp} \langle j_{s_\mp} | \frac{1}{\Omega}+ \frac{1}{2}  |
j_{s_\mp} \rangle + \nonumber \\ && 
B_{s_\pm,+} \langle j_{s_\mp} | \frac{1}{\Omega}+
\frac{1}{2}  | c_\ell \psi_+^{(0)}\rangle + 
 B_{s_\pm,-} \langle
j_{s_\mp} | \frac{1}{\Omega}+ \frac{1}{2}  | c_\ell \psi_-^{(0)}
\rangle = 0.
\end{eqnarray}
From symmetry considerations it follows that the terms with
$|\psi_\pm^{(0)}\rangle$ only contribute via $|p\rangle$ and by using
the definition of the transport values on the second term we obtain
\begin{equation}
B_{s_\pm,s_\mp} (s^\circ_\pm -  s^\circ_\mp) \langle s_\mp | s_\mp \rangle
+ (B_{s_\pm,+} +  B_{s_\pm,-}) \langle j_{s_\mp} | \frac{1}{\Omega}+
\frac{1}{2}  | c_\ell p \rangle = 0.
\end{equation}
Using the expressions for $B_{s_\pm,\sigma}$ and the fact that
$s^\circ_+$ and $s^\circ_-$ are different, it follows that
$B_{s_\pm,s_\mp}=0$.

The evaluation of the ${\cal N}$ now becomes straightforward. The last
term at the righthand site of Eq.~(\ref{Eq:N-mu}) will vanish because
of symmetry reasons. From the second term it can be observed that even
if we would not have chosen $B_{\mu\mu}=0$ as a normalization, this
coefficient would not contribute to the spectrum. For the viscosity we
immediately obtain ${\cal N}_\perp=0$. In the case of
$\mu=\sigma$ this results in
\begin{eqnarray}
{\cal N}_\sigma & = &\frac{\langle \rho | p \rangle^2}{2 \langle p
| p \rangle} \left( 1 + 2 \imath k \left[ B_{\sigma, -\sigma} +
B_{\sigma, s_+} \frac{\langle \rho | s_+ \rangle}{\langle \rho |
\psi_\sigma^{(0)} \rangle} + B_{\sigma, s_-} \frac{\langle \rho | s_-
\rangle}{\langle \rho | \psi_\sigma^{(0)} \rangle} \right] \right),
\nonumber \\
& = & \frac{ \langle \rho|\rho\rangle}{2 \gamma} \left(1 +
\frac{\imath \sigma k}{c_s} \left[ \Gamma + (\gamma-1) \chi\right]
\right),
\end{eqnarray}
where in the second line we have eliminated the viscosity. The
$|s_\pm\rangle$ lead to
\begin{eqnarray}
{\cal N}_{s_\pm} & = & \frac{(\gamma-1)^2 \langle \rho | p
\rangle^2}{2 \langle s_\pm | s_\pm \rangle} \left( 1 + 2 \imath k
\left[ B_{s_\pm,+} \frac{\langle \rho | \psi_+^{(0)} \rangle}{\langle
\rho | s_\pm^{(0)} \rangle} + B_{s_\pm,-} \frac{\langle \rho |
\psi_-^{(0)} \rangle}{\langle \rho | s_\pm\rangle} \right] \right),
\nonumber \\
& = & \frac{(\gamma-1) \langle \rho | \rho
\rangle}{2 \gamma} \left( 1 \pm \frac{\xi}{\sqrt{1 + \xi^2}} \right),
\end{eqnarray}
where the imaginary part cancels completely. 

\section{}
\label{Sec:appendixb}
In general one defines a transport coefficient via the decay of small
fluctuations with respect to the equilibrium distribution. If we take
$\delta f(\bm{k})$ to be such a fluctuation, we obtain for a single
timestep  
\begin{equation}
\label{Eq:transport2}
e^{z(\bm{k})} = \frac{ \langle \delta f | e^{-\imath \bm{k}\cdot
\bm{c}}(\bm{1}+\Omega) | \delta f \rangle}{ \langle \delta f
|\delta f \rangle},
\end{equation}
where the second order term in $k$ of $z(\bm{k})$ will be the transport
coefficient. 

Naturally these fluctuations can always be written in terms of the
eigenfunctions of the linearized Boltzmann operator
\begin{equation}
\label{Eq:signal}
|\delta f(\bm{k})\rangle = 
\sum_\mu \frac{\langle \psi_\mu | \delta f \rangle}{\langle \psi_\mu |
\psi_\mu \rangle} | \psi_\mu(\bm{k}) \rangle,
\end{equation}
where the coefficients on the right-hand site are independent of
$\bm{k}$ and therefore can be obtained from the $k=0$ limit
$\langle \psi_\mu | \delta f \rangle /\langle \psi_\mu |\psi_\mu
\rangle = \langle \psi_\mu^{(0)} | \delta f^{(0)} \rangle / \langle
\psi_\mu^{(0)} | \psi_\mu^{(0)} \rangle$. 
Combining this formal expansion with
Eq.~(\ref{Eq:transport2}) and using the results of the eigenvalue
problem (\ref{Eq:eigenproblem-right}) we obtain
\begin{equation}
e^{z(\bm{k})} = 
\frac{1} {\langle \delta f |\delta f \rangle} \sum_{\mu} \frac{\langle
\psi_\mu | \delta f \rangle}{\langle 
\psi_\mu | \psi_\mu \rangle} \langle \delta f | e^{z_\mu} |\psi_\mu
\rangle.
\end{equation}
If we restrict ourselves to fluctuations proportional to the
hydrodynamic modes, and thus assume the exponential fast decay of the
kinetic modes, the eigenvalues $e^{z_\mu}$ can be expanded in terms of
$\bm{k}$. This results in 
\begin{equation}
\label{Eq:Taylor}
e^{z(\bm{k})} = 1 + \sum_{\mu} \frac{\langle \psi_\mu | \delta f \rangle
\langle \delta f |\psi_\mu \rangle}{\langle \psi_\mu | \psi_\mu
\rangle \langle \delta f |\delta f \rangle} \left( \imath k
z_\mu^{(1)} - k^2 \left[ z_\mu^{(2)} + \frac{1}{2} (z_\mu^{(1)})^2
\right] \right) + {\cal O}(k^3).
\end{equation}
Note that the prefactors also depend on $\bm{k}$, but can be rewritten
as 
\begin{eqnarray}
& \frac{\langle \psi_\mu | \delta f \rangle \langle \delta f | \psi_\mu \rangle
}{\langle \psi_\mu | \psi_\mu \rangle \langle \delta f |\delta f \rangle} = 
\left( \frac{\langle \psi_\mu^{(0)} | \delta f^{(0)} \rangle}{\langle
\psi_\mu^{(0)} | \psi_\mu^{(0)} \rangle} \right)^2 \frac{\langle
\psi_\mu | \psi_\mu \rangle}{\langle \delta f |\delta f \rangle} =
\nonumber \\ &
\frac{\langle \psi_\mu^{(0)} | \delta f^{(0)} \rangle \langle \delta f^{(0)} |
\psi_\mu^{(0)} \rangle }{\langle \psi_\mu^{(0)} | \psi_\mu^{(0)}
\rangle \langle \delta f^{(0)} |\delta f^{(0)} \rangle} + {\cal O}(k^2).
\end{eqnarray}
This allows us to calculate the lowest order terms of $z(\bm{k})$ via
\begin{equation}
\label{Eq:order1}
z^{(1)} = \sum_\mu \frac{\langle \psi_\mu^{(0)} | \delta f^{(0)}
\rangle \langle \delta f^{(0)} | \psi_\mu^{(0)} \rangle }{\langle
\psi_\mu^{(0)} | \psi_\mu^{(0)} \rangle \langle \delta f^{(0)} |\delta f^{(0)}
\rangle} z_\mu^{(1)} 
\end{equation}
\begin{equation}
\label{Eq:order2}
z^{(2)} + \frac{1}{2} \left( z^{(1)} \right)^2 = \sum_\mu 
\frac{\langle \psi_\mu^{(0)} | \delta f^{(0)} \rangle \langle \delta f^{(0)} |
\psi_\mu^{(0)} \rangle }{\langle \psi_\mu^{(0)} | \psi_\mu^{(0)}
\rangle \langle \delta f^{(0)} |\delta f^{(0)} \rangle} \left( z_\mu^{(2)} +
\frac{1}{2} (z_\mu^{(1)})^2 \right) 
\end{equation}

It can easily be checked that in the case of a single eigenmode for
the fluctuation $\delta f$ these equations reduce to the normal
results, in particular Eq.~(\ref{Eq:transport}). In general, however,
such a formulation in terms of currents is not valid. The fact that it
nevertheless is consistent for Eqns.~(\ref{Eq:chi}) and (\ref{Eq:D})
is a consequence of the orthogonality relation (\ref{Eq:ortho-js})
between the currents of the two diffusive eigenmodes.

In the case of the different diffusions calculated in
Sec.~\ref{Sec:diffusion} we need to make use of these equations,
because the fluctuations under consideration are not eigenmodes. From
symmetry arguments it follows that in all those cases $z^{(1)}=0$, and
strictly speaking none of these fluctuations will therefore be
propagating. In the case of Eqns.~(\ref{Eq:diff-0}) and
(\ref{Eq:diffErnst}), however, the soundmodes will contribute to the
transport value (\ref{Eq:order2}) and spectra based on these
fluctuations will contain Brillouin peaks.

We finally like to mention that these results are only valid in the
limit $k\rightarrow 0$ and for small times, because if one considers
the decay for larger time intervals one would obtain
\begin{equation}
e^{z(\bm{k}) t} = \frac{ \langle \delta f | \left[ e^{-\imath \bm{k}\cdot
\bm{c}}(\bm{1}+\Omega) \right]^t| \delta f \rangle}{ \langle \delta f |\delta f
\rangle} = 
\sum_{\mu} 
\left( \frac{\langle \psi_\mu^{(0)} | \delta f^{(0)} \rangle}{\langle
\psi_\mu^{(0)} | \psi_\mu^{(0)} \rangle} \right)^2 \frac{\langle
\psi_\mu | \psi_\mu \rangle}{\langle \delta f |\delta f \rangle} e^{z_\mu t}.
\end{equation}
This results in a different behavior for short and long times in the
case one considers fluctuations that are not proportional to a single
eigenmode (See also Ref.~\cite{Mountain}).

\end{document}